\documentclass[reprint,amssymb,amsmath,aip,cha,frontmatterverbose]{revtex4-1}
\usepackage{graphicx}
\usepackage{dcolumn}
\usepackage{bm}
\usepackage{subfigure}

\begin{document}

\title{Two paradigmatic scenarios for inverse stochastic resonance}

\author{Iva Ba\v{c}i\'c}
\affiliation{Scientific Computing Laboratory, Center for the Study of Complex Systems,
Institute of Physics Belgrade, University of Belgrade, Pregrevica 118, 11080 Belgrade, Serbia}

\author{Igor Franovi\'c}
\email{franovic@ipb.ac.rs}
\affiliation{Scientific Computing Laboratory, Center for the Study of Complex Systems,
Institute of Physics Belgrade, University of Belgrade, Pregrevica 118, 11080 Belgrade, Serbia}

\date{\today}% It is always \today

\begin{abstract}
Inverse stochastic resonance comprises a nonlinear response of an oscillatory system to noise where the frequency of noise-perturbed oscillations becomes minimal at an intermediate noise level. We demonstrate two generic scenarios for inverse stochastic resonance by considering a paradigmatic model of two adaptively coupled stochastic active rotators whose local dynamics is close to a bifurcation threshold. In the first scenario, shown for the two rotators in the excitable regime, inverse stochastic resonance emerges due to a biased switching between the oscillatory and the quasi-stationary metastable states derived from the attractors of the noiseless system. In the second scenario, illustrated for the rotators in the oscillatory regime, inverse stochastic resonance arises due to a trapping effect associated with a noise-enhanced stability of an unstable fixed point. The details of the mechanisms behind the resonant effect are explained in terms of slow-fast analysis of the corresponding noiseless systems.
\end{abstract}

\maketitle

\begin{quotation}
The effects of noise may generically be classified into two groups: on the one hand, the noise may enhance or suppress certain features of deterministic dynamics by acting on the system states in an inhomogeneous fashion, while on the other hand, it may give rise to novel forms of behavior, associated with crossing of thresholds and separatrices, or to a stability of deterministically unstable states. The constructive role of noise has been evinced in a wide range of real-world applications, from neural networks and chemical reactions to lasers and electronic circuits. The classical examples of stochastic facilitation concern the resonant phenomena, including stochastic resonance, where noise of appropriate intensity may induce oscillations in bistable systems that are preferentially locked to a weak periodic forcing, and coherence resonance, where an intermediate level of noise may trigger coherent oscillations in excitable systems. Recently, a novel form of nonlinear response to noise, called inverse stochastic resonance, has been discovered while studying individual neural oscillators and models of neuronal populations. It has come to light that noise may reduce the intrinsic spiking frequency of neuronal oscillators, transforming the tonic firing into a bursting-like activity or even quenching the oscillations. Within the present study, we demonstrate two paradigmatic mechanisms of inverse stochastic resonance, one based on biased switching between the metastable states, and the other associated with a noise-enhanced stability of an unstable fixed point. We show that the effect is robust, in a sense that it may emerge in coupled excitable and coupled oscillatory systems, and both in cases of Type I and Type II oscillators.
\end{quotation}

Noise in excitable or multistable systems may fundamentally change their deterministic dynamics, giving rise to qualitatively novel forms of behavior, associated with crossing of thresholds and separatrices, or stabilization of certain unstable structures \cite{LGNS04,FM18}. The emergent dynamics may involve noise-induced oscillations and stochastic bursting \cite{NR02,RS17,ZP18}, switching between metastable states \cite{BYWF18,FK18} or noise-enhanced stability of metastable and unstable states \cite{FSB05,MS96,CPS93,ADS03,AVS10}, to name but a few. In neuronal systems, the phenomena reflecting the constructive role of noise are collected under the notion of stochastic facilitation \cite{MW11,SM13,DR12}, which mainly comprises the resonant effects. The most prominent examples concern coherence resonance \cite{PK97,LSG99,MNV01,ZFVS13,SZAS16}, where the regularity of noise-induced oscillations becomes maximal at a preferred noise level, and stochastic resonance \cite{GHJM98,MW11}, where the sensitivity of a system to a subthreshold periodic stimulation becomes maximal at an intermediate noise level. Recent studies on the impact of noise in neuronal oscillators have revealed that the noise may also give rise to an inhibitory effect, which consists in reducing the intrinsic spiking frequency, such that it becomes minimal at an intermediate noise intensity \cite{SM13,GJT08,TJG09,UCOB13,U13,UTSOB17,UBT17,BRHGR16,BKNPF18,FOW18}. This effect has been called inverse stochastic resonance (ISR), but in contrast to  stochastic resonance, it concerns autonomous rather than periodically driven systems. Apart from reports in models of neurons and neuronal populations, ISR has recently been evinced for cerebellar Purkinje cells \emph{in-vitro} \cite{BRHGR16}, having shown how the lifetimes of the so-called UP states with elevated spiking activity and the DOWN states of relative quiescence \cite{FK18,HMS12,VH13,FK16} depend on the noise intensity.

The studies of the mechanism behind ISR have so far mostly been focused on Type II neural oscillators with bistable dynamics poised close to a subcritical Hopf bifurcation \cite{TJG09,UCOB13,SM13,U13}, considering Hodgkin-Huxley and Morris-Lecar models. Under the influence of noise, such systems exhibit switching between the two metastable states, derived from the periodic and the stationary attractor of the deterministic dynamics. At an intermediate noise level, one observes that the switching rates become strongly asymmetric, with the system spending substantially more time in a quasi-stationary state. This is reflected in a characteristic non-monotone dependence of the spiking frequency on noise, which is a hallmark of ISR.

Nevertheless, a number of important issues on the mechanism giving rise to ISR have remained unresolved. In particular, is the effect dependent on the type of neuronal excitability? Also, can there be more than a single mechanism of ISR? And finally, how does the effect depend on the form of couplings and whether it can be robust for adaptively changing couplings, typical for neuronal systems?

To address these issues, we invoke a simple, yet paradigmatic model that combines the three typical ingredients of neuronal dynamics, including excitability, noise and coupling plasticity. In particular, we consider a system of two identical, adaptively coupled active rotators \cite{BYWF18,BKNPF18,KPLS14} influenced by independent Gaussian white noise sources
\begin{align}
\dot{\varphi_i}&= I_0 - \sin{\varphi_i} + \kappa_i \sin{(\varphi_j-\varphi_i)} + \sqrt{D}\xi_i(t) \nonumber \\
\dot{\kappa_i}&= \varepsilon (-\kappa_i + \sin(\varphi_j - \varphi_i + \beta)). \label{eq1}
\end{align}
The indices $i,j\in\{1,2\}, i \neq j$ denote the particular units, described by the respective phases $\{\varphi_{1},\varphi_{2}\}\in S^{1}$, which constitute the fast variables, and the slowly varying coupling weights $\{\kappa_{1},\kappa_{2}\}\in \mathcal{R}$. The scale separation between the characteristic timescales is set by the small parameter $\varepsilon \ll 1$ that defines the adaptivity rate. The local dynamics is controlled by the excitability parameter $I_0$, such that the saddle-node of infinite period (SNIPER) bifurcation at $I_0=1$ mediates the transition between the excitable ($I_0 \lesssim 1$) and the oscillatory regime ($I_0>1$). The excitable units may still exhibit oscillations, induced either by the action of the coupling (\emph{emergent} oscillations) and/or evoked by the stochastic terms (noise-induced oscillations). The noiseless coupled system \eqref{eq1} is invariant with respect to exchange of the units' indices, such that all the stationary or the periodic solutions always appear in pairs connected by the $Z_2$ symmetry. Given the similarity between the active rotators and the theta neurons, which also conform to Type I excitability, system \eqref{eq1} may be considered qualitatively analogous to a motif of two adaptively coupled neurons \cite{SK04}, influenced by an external bias current $I_0$ and the synaptic noise. Adaptivity is modeled in terms of phase-dependent plasticity \cite{MLHBT07,AA09,AA11} of coupling weights, having the modality of the plasticity rule adjusted by the parameter $\beta$. This form of plasticity has already been shown capable of qualitatively reproducing the features of some well-known neuronal plasticity rules \cite{AA09,AA11}. In particular, for $\beta=3\pi/2$, one recovers Hebbian-like learning \cite{H49}, where the synaptic potentiation promotes phase synchronization, while for $\beta=\pi$, adaptation acts similar to spike-timing-dependent plasticity (STDP) \cite{SMA00,FD02,WGNB05,MDG08,PYT13}, whose typical form \cite{MDG08,LPTY16} favors a causal relationship between the pre- and post-synaptic neuron firing times \cite{AA09,AA11}.

\vspace{-0.3cm}
\section{Inverse stochastic resonance due to a biased switching} \label{sec:m1}

The first generic scenario for ISR we demonstrate is based on \emph{biased switching} between the metastable states associated with coexisting stationary and periodic attractors of the corresponding deterministic system. As an example, we consider the noise-induced reduction of frequency of emergent oscillations on a motif of two adaptively coupled stochastic active rotators with excitable local dynamics ($I_0=0.95$). To elucidate the mechanism behind the effect, we first summarize the details of the noise-free dynamics, and then address the switching behavior. A complete bifurcation analysis of the noiseless version of \eqref{eq1} with excitable local dynamics has been carried out in \cite{BYWF18,BKNPF18}, having shown (i) how the number and stability of the fixed points depends on the plasticity rule, characterized by $\beta$, as well as (ii) how the interplay between $\beta$ and the adaptivity rate, controlled by the small parameter $\varepsilon$, gives rise to limit cycle attractors. Our focus is on the interval $\beta\in(3.298,4.495)$, which approximately interpolates between the limiting cases of Hebbian-like and STDP-like plasticity rules. There, the system exhibits two stable equilibria born from the symmetry-breaking pitchfork bifurcation, and has four additional unstable fixed points. For the particular case $\beta=4.2$ analyzed below, the two stable equilibria, given by EQ1$:=(\varphi_1^*,\varphi_2^*,\kappa_1^*,\kappa_2^*)=(1.2757,0.2127,-0.0078,-0.8456)$ and EQ2$:=(\varphi_1^*,\varphi_2^*,\kappa_1^*,\kappa_2^*)=(0.2127,1.2757,-0.8456,-0.0078)$, have been shown to manifest excitable behavior \cite{BYWF18}.

\begin{figure*}
\centering
\includegraphics[scale=0.42]{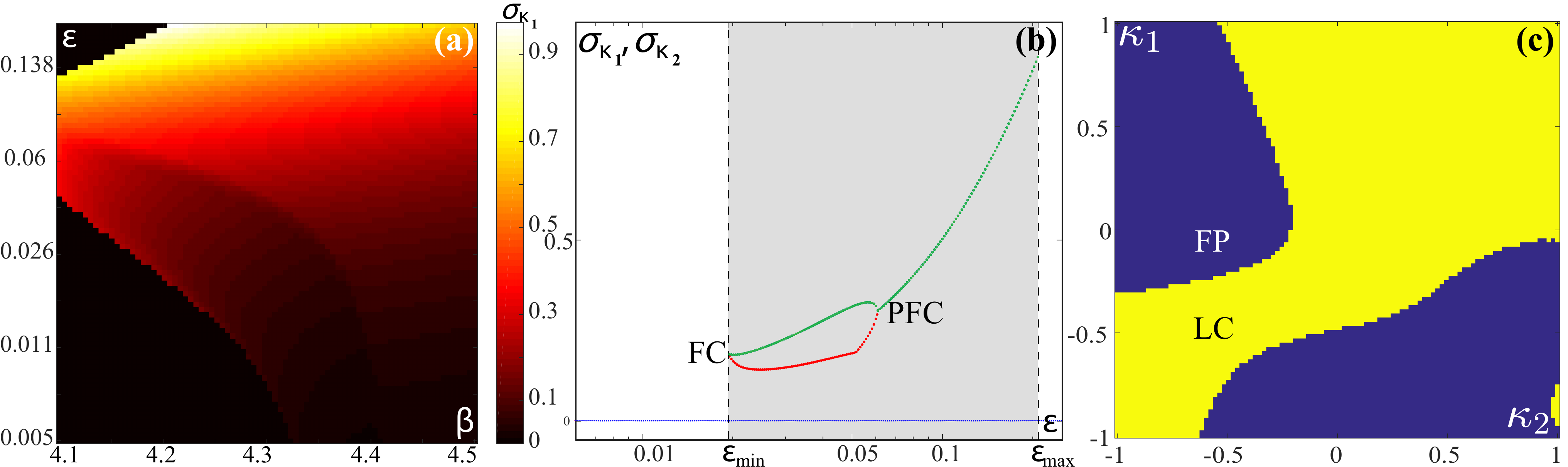}
\caption{Emergent oscillations in \eqref{eq1} for $I_0=0.95,D=0$. (a) Variation $\sigma_{\kappa_1}$ of the coupling weight $\kappa_1$ in the $(\beta,\varepsilon)$ plane. (b) Dependencies $\sigma_{\kappa_i}(\varepsilon),i\in\{1,2\}$ for the representative stationary (blue) and
oscillatory solution (red and green refer to the two units) at fixed $\beta=4.2$. Shading indicates the $\varepsilon$ interval that supports multistability between the two symmetry-related stable equilibria and the limit cycle attractor(s). FC and PFC denote the $\varepsilon$ values where the fold of cycles and pitchfork of cycles occur. (c) Basins of stability of the stationary (FP, blue) and oscillatory solutions (LC, yellow) in the $(\kappa_1,\kappa_2)$ plane, obtained by fixing the initial phases to $(\varphi_1,\varphi_2)=(1.32,0.58)$. The remaining parameters are $\beta=4.2,\varepsilon=0.1$.} \label{fig1}
\vspace{-0.6cm}
\end{figure*}
The onset of emergent oscillations, as well as the coexistence between the stable stationary and periodic solutions in the noiseless version of \eqref{eq1}, are illustrated in Fig. \ref{fig1}. The maximal stability region of the two $Z_2$ symmetry-related periodic solutions is indicated in Figure \ref{fig1}(a), which shows the variation of the $\kappa_1$ variable, $\sigma_{\kappa_1}=\max(\kappa_1(t))-\min(\kappa_1(t))$, in the $(\beta,\varepsilon)$ parameter plane. The scan was performed by the method of numerical continuation starting from a stable periodic solution, such that the initial conditions for an incremented parameter value are given by the final state obtained for the previous iteration step. One finds that for a given $\beta$, there exists an interval $\varepsilon\in(\varepsilon_{min},\varepsilon_{max})$ of intermediate scale separation ratios supporting the oscillations, cf. the highlighted region in Fig. \ref{fig1}(b). In particular, the two $Z_2$-symmetry related branches of stable periodic solutions emanate from the fold of cycles bifurcations, denoted by FC in Fig. \ref{fig1}(b), such that the associated threshold scale separation $\varepsilon_{min}(\beta)$ decreases with $\beta$. The two branches of oscillatory solutions merge around $\varepsilon\approx 0.06$, where the system undergoes an inverse pitchfork bifurcation of limit cycles (PFC). The incipient stable limit cycle acquires the anti-phase space-time symmetry $\varphi_1(t)=\varphi_2(t+T_{osc}/2),\kappa_1(t)=\kappa_2(t+T_{osc}/2)$, with $T_{osc}$ denoting the oscillation period \cite{BYWF18}. An example illustrating the basins of stability of stationary and oscillatory solutions for $\varepsilon=0.1$, obtained by fixing the initial values of phases and varying the initial coupling weights within the range $\kappa_{i,ini}\in(-1,1)$, is shown in Figure \ref{fig1}(c). In the presence of noise, the coexisting attractors of the deterministic system turn to metastable states, which are connected by the noise-induced switching.

\begin{figure*}
\centering
\includegraphics[scale=0.16]{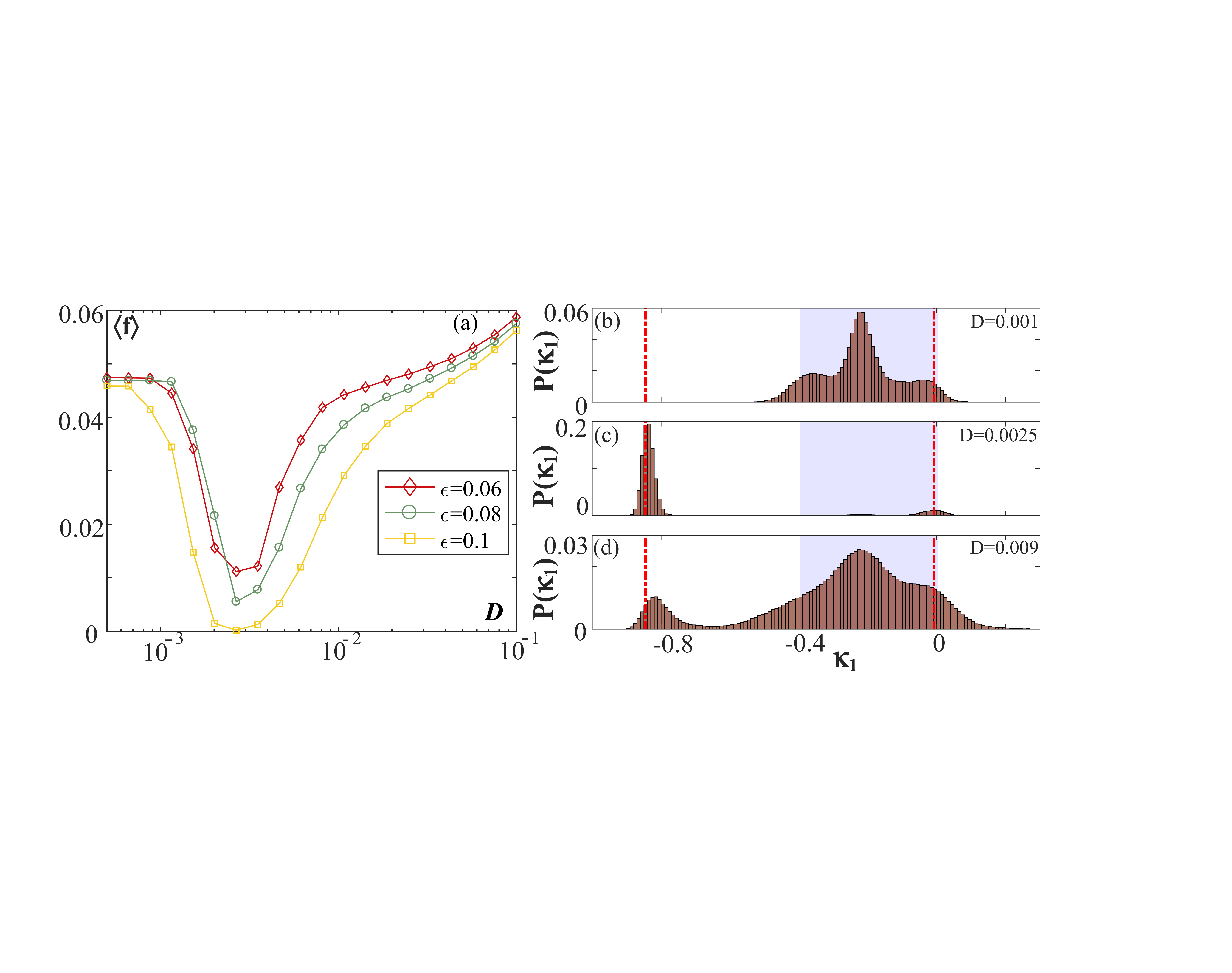}
\caption{(a) Dependencies of the mean oscillation frequency on noise for scale separation $\varepsilon=0.06$ (diamonds), $\varepsilon=0.08$ (circles) and $\varepsilon=0.1$ (squares), obtained for fixed $I_0=0.95,\beta=4.2$. Averaging has been performed over an ensemble of 100 different stochastic realizations. (b)-(d) show the stationary distributions $P(\kappa_1)$ below ($D=0.001$), at ($D=0.0025$) and above ($D=0.009$) the resonant noise intensity for $\varepsilon=0.1$. The dash-dotted lines denote the $\kappa_1$ levels associated with the two stable equilibria, $\kappa_1^{*}(EQ1)$ and $\kappa_1^{*}(EQ2)$, while the blue shaded interval indicates the variation $\sigma_{\kappa_1}$ of the unique stable periodic solution.} \label{fig2}
\vspace{-0.4cm}
\end{figure*}
Inverse stochastic resonance manifests itself as the noise-mediated suppression of oscillations, whereby the frequency of noise-perturbed oscillations becomes minimal at an intermediate noise level. For the motif of two adaptively coupled excitable active rotators, such characteristic non-monotone dependence on noise is generically found for intermediate adaptivity rates supporting multistability between the stationary and the oscillatory solutions. A family of curves illustrating the dependence of the oscillation frequency on noise variance $\langle f\rangle(D)$ for a set of different $\varepsilon$ values is shown in Fig. \ref{fig2}(a). The angular brackets $\langle \cdot\rangle$ refer to averaging over an ensemble of a 100 different stochastic realizations, having fixed a set of initial conditions within the basin of attraction of the limit cycle attractor. Nonetheless, qualitatively analogous results are recovered if for each realization of the stochastic process, one selects a set of random initial conditions lying within the stability basin of a periodic solution. In \cite{BKNPF18}, we have shown that the noise-induced switching gives rise to a bursting-like behavior, where the spiking is interspersed by the quiescent episodes which correspond to the system residing in the vicinity of the quasi-stationary metastable states. Such episodes become prevalent at the noise levels around the minimum of $\langle f\rangle(D)$. For weaker noise $D\lesssim 10^{-3}$, the frequency of emergent oscillations remains close to the deterministic one, whereas for a much stronger noise, it increases above that of unperturbed oscillations. One observes that the suppression effect of noise depends on the adaptivity rate, such that it is enhanced for faster adaptivity, see \cite{BKNPF18} for a more detailed analysis. In order to illustrate how the ISR effect is reflected at the level of the dynamics of coupling weights, in Fig. \ref{fig2}(b)-(d) are shown the stationary distributions $P(\kappa_1)$ for the noise levels below, at and above the resonant level. To provide a reference to the deterministic case, we have denoted by the dash-dotted lines the weight levels associated with the two equilibria EQ1 and EQ2, while the blue shading indicates the variation $\sigma_{\kappa}$ of the stable limit cycle. Note that the stable periodic solution is unique because for the considered $\varepsilon$ value, the deterministic system lies above the pitchfork of cycles bifurcation, cf. PFC in Fig. \ref{fig1}(b). The stationary distribution $P(\kappa_1)$ at the resonant noise expectedly shows a pronounced peak at one of the quasi-stationary states, while the distributions below or above the resonant noise level indicate a high occupancy of the oscillatory metastable state.

\begin{figure*}
\centering
\includegraphics[scale=0.65]{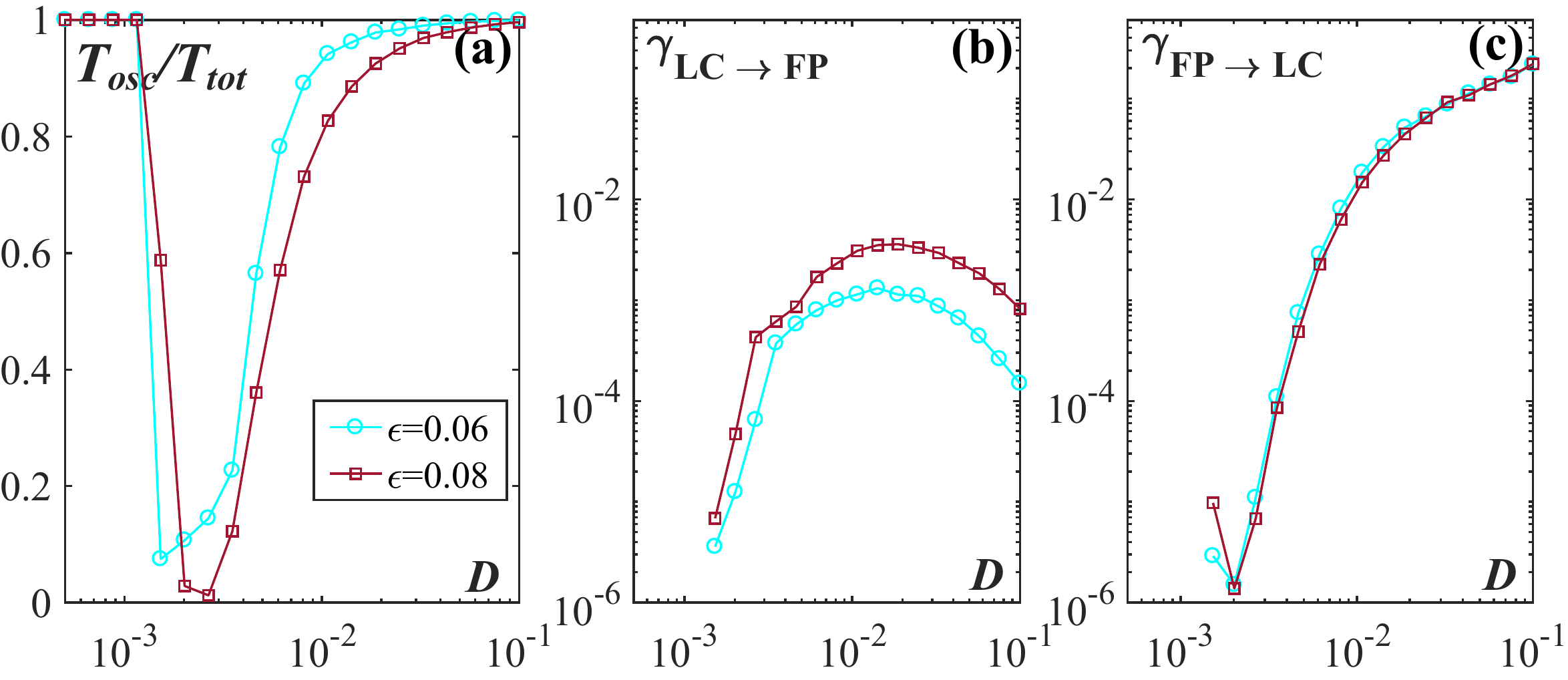}
\caption{(a) Fraction of the time spent at the oscillatory metastable state $T_{osc}/T_{tot}$ as a function of noise for
$\varepsilon=0.06$ (circles) and $\varepsilon=0.08$ (squares). (b) and (c) Numerically estimated transition rates from the
oscillatory to the quasi-stationary metastable states, $\gamma_{LC\rightarrow FP}(D)$ and \emph{vice versa}, $\gamma_{FP\rightarrow LC}(D)$.
The remaining parameters are $I_0=0.95,\beta=4.2$.}\label{fig3}
\vspace{-0.5cm}
\end{figure*}
In order to elucidate the mechanism behind ISR, we have calculated how the fraction of the total time spent at the oscillatory metastable states, $T_{osc}/T_{tot}$, changes with noise. In terms of numerical experiments, the quasi-stationary and the oscillatory metastable states can readily be distinguished by considering the corresponding $\kappa_i(t)$ series, using the fact that the typical distance $|\kappa_1(t)-\kappa_2(t)|$ is much larger for the quasi-stationary than the oscillatory solutions. This has allowed us to employ a simple threshold method to identify the particular system's states and trace the associated transitions. Figure \ref{fig3}(a) indicates a non-monotone dependence of $T_{osc}/T_{tot}(D)$, implying that the switching process around the resonant noise level becomes strongly biased toward the quasi-stationary state, even more so for a faster adaptivity. The biased switching is facilitated by the geometry of the phase space, featuring an asymmetrical structure with respect to the separatrix between the coexisting attractors, such that the limit cycle lies much closer to the separatrix than the stationary states.

The nonlinear response to noise may be understood in terms of the competition between the transition processes from and to the limit cycle attractor. These processes are characterized by the transition rates from the stability basin of the limit cycle attractor to that of the stationary states $\gamma_{LC\rightarrow FP}$ and \emph{vice versa}, $\gamma_{FP\rightarrow LC}$, which are numerically estimated as the reciprocal values of the corresponding mean first-passage times \cite{HTB90}. In Fig. \ref{fig3}(b)-(c) is illustrated the qualitative distinction between the noise-dependencies of the transition rates: while $\gamma_{LC\rightarrow FP}$ displays a maximum at the resonant noise level, $\gamma_{FP\rightarrow LC}$ just increases monotonously with noise. For small noise $D\lesssim 10^{-3}$, one observes virtually no switches to quasi-stationary state, as evinced by the fact that the corresponding oscillation frequency is identical to the deterministic one. For increasing noise, the competition between the two processes is resolved in such a way that at an intermediate/large noise, the impact of $\gamma_{LC\rightarrow FP}$/$\gamma_{FP\rightarrow LC}$ becomes prevalent. The large values of
$\gamma_{FP\rightarrow LC}$ found for quite strong noise $D\gtrsim 0.04$ reflect the point that the system there spends most of the time in the oscillatory metastable state, making only quite short excursions to the quasi-stationary state.

Though ISR is most pronounced for intermediate $\varepsilon$, it turns out that an additional subtlety in the mechanism of biased switching may be explained by employing the singular perturbation theory to the noiseless version of \eqref{eq1}. In particular, by combining the critical manifold theory \cite{K15} and the \emph{averaging} approach \cite{S08}, one may demonstrate the \emph{facilitatory role of plasticity} in enhancing the resonant effect, showing that the adaptation drives the fast flow toward the parameter region where the stationary state is a focus rather than a node \cite{BKNPF18}. The response to noise in multiple timescale systems has already been indicated to qualitatively depend on the character of the stationary states, yielding fundamentally different scaling regimes with respect to noise variance and the scale-separation ratio \cite{BG06,LL09,TW15}. Intuitively, one expects that the resonant effects should be associated with the quasi-stationary states derived from the focuses rather than the nodes \cite{BG06}, because the local dynamics then involves an eigenfrequency.

\begin{figure*}
\centering
\includegraphics[scale=0.31]{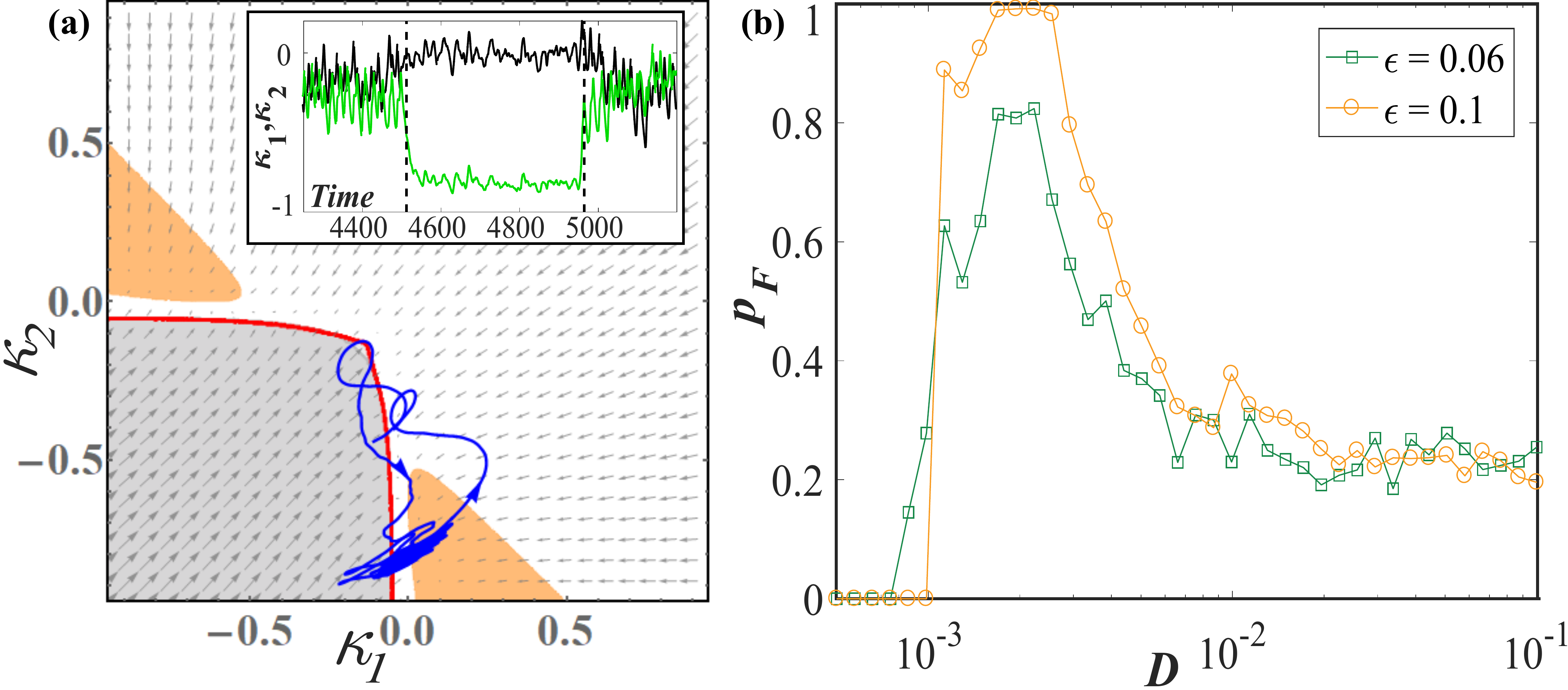}
\caption{(a) Fast-slow analysis of \eqref{eq1} for $I_0=0.95,D=0$. The fast flow exhibits a periodic attractor (grey shaded region) and a stable equilibrium (white region), with two branches of SNIPER bifurcations (red lines) outlining the boundary between them. The arrows indicate the vector fields corresponding to the stable sheets of the slow flow. The inset shows $\kappa_i(t)$ series corresponding to a switching episode from the oscillatory to the stationary state and back, obtained for $\varepsilon=0.06,\beta=4.2$. The corresponding $(\kappa_1(t),\kappa_2(t))$ orbit is indicated by the blue line. Within the two orange regions, the two stable equilibria are focuses rather than the nodes. (b) Conditional probability $p_F(D)$ of having the crossing of SNIPER bifurcation followed by a visit to the orange shaded region from (a), obtained for $\varepsilon=0.06$ (squares) and $\varepsilon=0.1$ (circles).} \label{fig4}
\vspace{-0.6cm}
\end{figure*}

The fast-slow analysis of \eqref{eq1} for $I_0=0.95$ has been carried out in detail in \cite{BYWF18,BKNPF18}, such that here we only summarize the main results concerning the associated layer and reduced problems \cite{K15}. Within the layer problem, the fast flow dynamics
\begin{align}
\dot{\varphi_1}&= I_0 - \sin{\varphi_1} + \kappa_1 \sin{(\varphi_2-\varphi_1)} \nonumber \\
\dot{\varphi_2}&= I_0 - \sin{\varphi_2} + \kappa_2 \sin{(\varphi_1-\varphi_2)}, \label{eq2}
\end{align}
is considered by treating the slow variables $\kappa_1,\kappa_2\in[-1,1]$ as additional system parameters. Depending on $\kappa_1$ and $\kappa_2$, the fast flow dynamics is found to be almost always \emph{monostable}, exhibiting either a stable equilibrium or a limit cycle attractor, apart from a small region of bistability between the two \cite{BYWF18,BKNPF18}. The maximal stability region of the oscillatory regime, encompassing both the domain where the oscillatory solution is monostable and where it coexists with a stable equilibrium, is indicated by the gray shading in Fig. \ref{fig4}(a). The latter has been determined by the method of numerical continuation, starting from a periodic solution. The thick red lines outlining the region’s boundaries correspond to the two branches of SNIPER bifurcations \cite{BYWF18}.  Note that for each periodic solution above the main diagonal $\kappa_1=\kappa_2$, there exists a $Z_2$ symmetry-related counterpart below the diagonal.

By averaging over the different attractors of the fast flow dynamics, we have obtained multiple stable sheets of the slow flow \cite{S08}. The explicit procedure consists in determining the time average $\langle \varphi_2-\varphi_1\rangle_t=h(\kappa_1,\kappa_2)$ by iterating \eqref{eq2} for each fixed set $(\kappa_1,\kappa_2)$ \cite{BYWF18,S08}, and then substituting these averages into the equations of the slow flow
\begin{align}
\kappa_1^{'}&=[-\kappa_1+\sin(h(\kappa_1,\kappa_2)+\beta)]\nonumber \\
\kappa_2^{'}&= [-\kappa_2+\sin(-h(\kappa_1,\kappa_2)+\beta)], \label{eq3}
\end{align}
where the prime refers to a derivative over the rescaled time variable $T:=t/\varepsilon$. The arrows in Fig. \ref{fig4}(a) show the vector fields on the two stable sheets of the slow flow \eqref{eq3} associated with the stationary and the periodic attractors of the fast flow.

The performed fast-slow analysis has allowed us to gain a deeper insight into the facilitatory role of adaptivity within the ISR. In particular, in the inset of Fig. \ref{fig4}(a) are extracted the time series $(\kappa_1(t),\kappa_2(t))$ which (from left to right) illustrate the switching episode from an oscillatory to the quasi-stationary metastable state. The triggering/termination of this switching event is associated with an inverse/direct SNIPER bifurcation of the fast flow. Note that for $(\kappa_1,\kappa_2)$ values immediately after the inverse SNIPER bifurcation, the stable equilibrium of the fast flow is a node. Nevertheless, for the noise levels corresponding to the most pronounced ISR effect, the coupling dynamics guides the system into the triangular orange-shaded regions in Fig. \ref{fig4}(a), where the equilibrium is a stable focus rather than a node. We have verified that this feature is a hallmark of ISR by numerically calculating the conditional probability $p_F$ that the events of crossing the SNIPER bifurcation are followed by the system's orbit visiting the $(\kappa_1,\kappa_2)$ regions with a focus equilibrium. The $p_F(D)$ dependencies for two characteristic $\varepsilon$ values in Fig. \ref{fig4}(b) indeed show a maximum for the resonant noise levels, corresponding to the minima of the frequency dependencies in Fig. \ref{fig2}(a). The local dynamics around the focus gives rise to a \emph{trapping} effect, such that the phase variables remain for a longer time in the associated quasi-stationary states than in case where the metastable states derive from the nodes of the fast flow. Small noise below the resonant values is insufficient to drive the system to the regions featuring focal equilibria, whereas for too strong noise, the stochastic fluctuations completely take over, washing out the quasi-stationary regime. The trapping effect is enhanced for the faster adaptivity rate, as evinced by the fact that the curve $p_F(D)$ for $\varepsilon=0.1$ lies above the one for $\varepsilon=0.06$.

\vspace{-0.4cm}
\section{Inverse stochastic resonance due to a trapping effect}
\vspace{-0.2cm}

As the second paradigmatic scenario for ISR, we consider the case where the oscillation frequency is reduced due to a noise-induced trapping in the vicinity of an \emph{unstable} fixed point of the noiseless system. Such a trapping effect may be interpreted as an example of the phenomenon of \emph{noise-enhanced stability} of an unstable fixed point \cite{FSB05,MS96,CPS93,ADS03,AVS10}. This mechanism is distinct from the one based on biased switching, because there the quasi-stationary states derive from the \emph{stable} equilibria of the noise-free system, such that the noise gives rise to crossing over the separatrix between the oscillatory and the quiescent regime. Nevertheless, in the scenario below, noise induces "tunneling" through the bifurcation threshold, temporarily stabilizing an unstable fixed point of the deterministic system.

\begin{figure}
\centering
\includegraphics[scale=0.66]{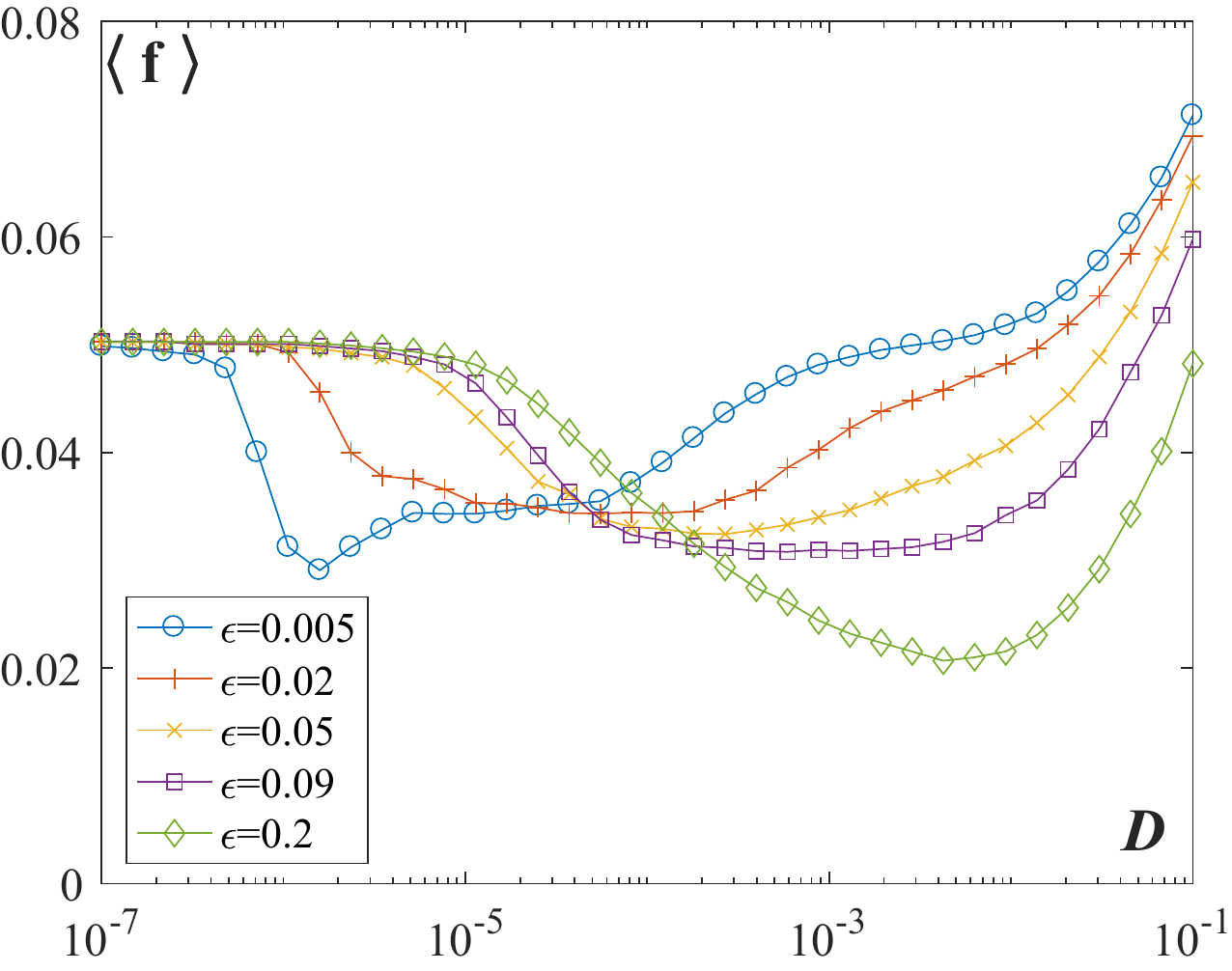}
\caption{Family of dependencies $\langle f\rangle(D)$ for scale separations $\varepsilon\in\{0.005,0.02,0.05,0.09,0.2\}$ at fixed $I_0=1.05,\beta=\pi$. Stochastic averaging has involved an ensemble of 100 different process realizations.} \label{fig5}
\vspace{-0.3cm}
\end{figure}
In particular, we study an example of a system \eqref{eq1} comprised of two adaptively coupled active rotators in the \emph{oscillatory}, rather than the excitable regime, setting the parameter $I_0=1.05$ close to a bifurcation threshold.  The plasticity parameter is fixed to $\beta=\pi$, such that the modality of the phase-dependent adaptivity resembles the STDP rule in neuronal systems. One finds that this system exhibits a characteristic non-monotone response to noise, with the oscillation frequency of the phases $\langle f\rangle$ displaying a minimum at an intermediate noise level, see Fig. \ref{fig5}. In contrast to the mechanism described in Sec. \ref{sec:m1}, the onset of ISR here does not qualitatively depend on the adaptivity rate. One only finds a quantitative dependence of the system’s nonlinear response to noise on $\varepsilon$, in a sense that the resonant noise level shifts to larger values with increasing $\varepsilon$.  Our exhausitive numerical simulations indicate that the ISR effect persists for slow adaptivity rates, cf. the example of the $\langle f(D)\rangle$  for $\varepsilon=0.005$ in Fig. \ref{fig5}, and the results of the fast-slow analysis below will further show that all the ingredients required for the ISR effect remain in the singular perturbation limit $\varepsilon \rightarrow 0$. The persistence of the ISR effect has also been numerically confirmed for faster adaptivity rates $\varepsilon \sim 0.1$. In this case, we have observed that the minima of the $\langle f(D)\rangle$ curves become deeper with $\varepsilon$, suggesting that the ISR becomes more pronounced for higher adaptivity rates.

To elucidate the mechanism behind ISR, we again perform the fast-slow analysis of the corresponding noise-free system. Prior to this, we briefly summarize the results of the numerical bifurcation analysis for the noiseless system in the case of finite scale separation. First note that selecting a particular plasticity rule $\beta=\pi$ confines the dynamics of the couplings to a symmetry invariant subspace $\kappa_1(t)=-\kappa_2(t) \equiv\kappa(t)$. Due to this, the noiseless version of the original system \eqref{eq1} can be reduced to a three-dimensional form
\begin{align}
\dot{\varphi_1}&= I_0 - \sin{\varphi_1} + \kappa \sin{(\varphi_2-\varphi_1)} \nonumber \\
\dot{\varphi_2}&= I_0 - \sin{\varphi_2} + \kappa \sin{(\varphi_2-\varphi_1)} \nonumber \\
\dot{\kappa}&= \varepsilon (-\kappa - \sin(\varphi_2 - \varphi_1)).    \label{eq4}
\end{align}
By numerically solving the eigenvalue problem, we have verified that \eqref{eq4} possesses no stable fixed points, but rather a pair of saddle nodes and a pair of saddle focuses. Also, we have determined that the maximal real part of the eigenvalues of the focuses displays a power-law dependence on the scale separation, tending to zero in the singular limit $\varepsilon\rightarrow 0$. Concerning the oscillatory states, our numerical experiments show that \eqref{eq4} exhibits multistability between three periodic solutions, whereby two of them are characterized by the non-zero couplings and a constant phase-shift between the fast variables, whereas the third solution corresponds to a case of effectively uncoupled units $(\kappa(t)=0)$ and the fast variables synchronized in-phase.

A deeper understanding of the ingredients relevant for the trapping mechanism can be gained within the framework of fast-slow analysis, considering the layer problem
\begin{align}
\dot{\varphi_1}&= I_0 - \sin{\varphi_1} + \kappa \sin{(\varphi_2-\varphi_1)} \nonumber \\
\dot{\varphi_2}&= I_0 - \sin{\varphi_2} + \kappa \sin{(\varphi_2-\varphi_1)}. \label{eq5}
\end{align}
Treating $\kappa\in[-1,1]$ as an additional system parameter, we first look for the stationary and periodic attractors of the fast flow. It is convenient to apply the coordinate transformation $(\varphi_1, \varphi_2) \mapsto (\Phi, \delta \varphi) = (\frac{\varphi_1+\varphi_2}{2}, \frac{\varphi_1-\varphi_2}{2})$, rewriting \eqref{eq5} as
\begin{align}
\delta \dot{\varphi} &= - \sin \delta \varphi \cos \Phi \nonumber \\
\dot{\Phi} &= I_0 - \cos \delta \varphi  (\sin \Phi +2 \kappa \sin \delta \varphi).  \label{eq6}
\end{align}
From the second equation, one readily finds that the fast flow cannot possess any fixed points on the synchronization manifold $\delta \varphi=0$ because $I_0>1$, such that the stationary solutions derive only from the condition $\cos \Phi=0$. A numerical analysis shows that, depending on $\kappa$, the fast flow for $I_0\gtrsim 1$ can exhibit two or no fixed points. For the particular value $I_0=1.05$, one finds that two fixed points, namely a saddle and a \emph{center}, exist within the interval $\kappa\in[-0.1674,0.1674]$. The appearance of a center point is associated with the time-reversal symmetry of the fast flow \eqref{eq5}. Indeed, one may show that the fast flow is invariant to a symmetry-preserving map $R$ of the form
\begin{equation}
R=
\begin{cases}
\varphi_1 \rightarrow \pi - \varphi_2,\\
\varphi_2 \rightarrow \pi - \varphi_1,\\
t \rightarrow -t
\end{cases} \label{eq7}
\end{equation}
Note that in case of the finite scale separation, the counterpart of the center point of the fast flow is a weakly unstable focus of the complete system \eqref{eq4}.

\begin{figure*}
\centering
\includegraphics[scale=0.35]{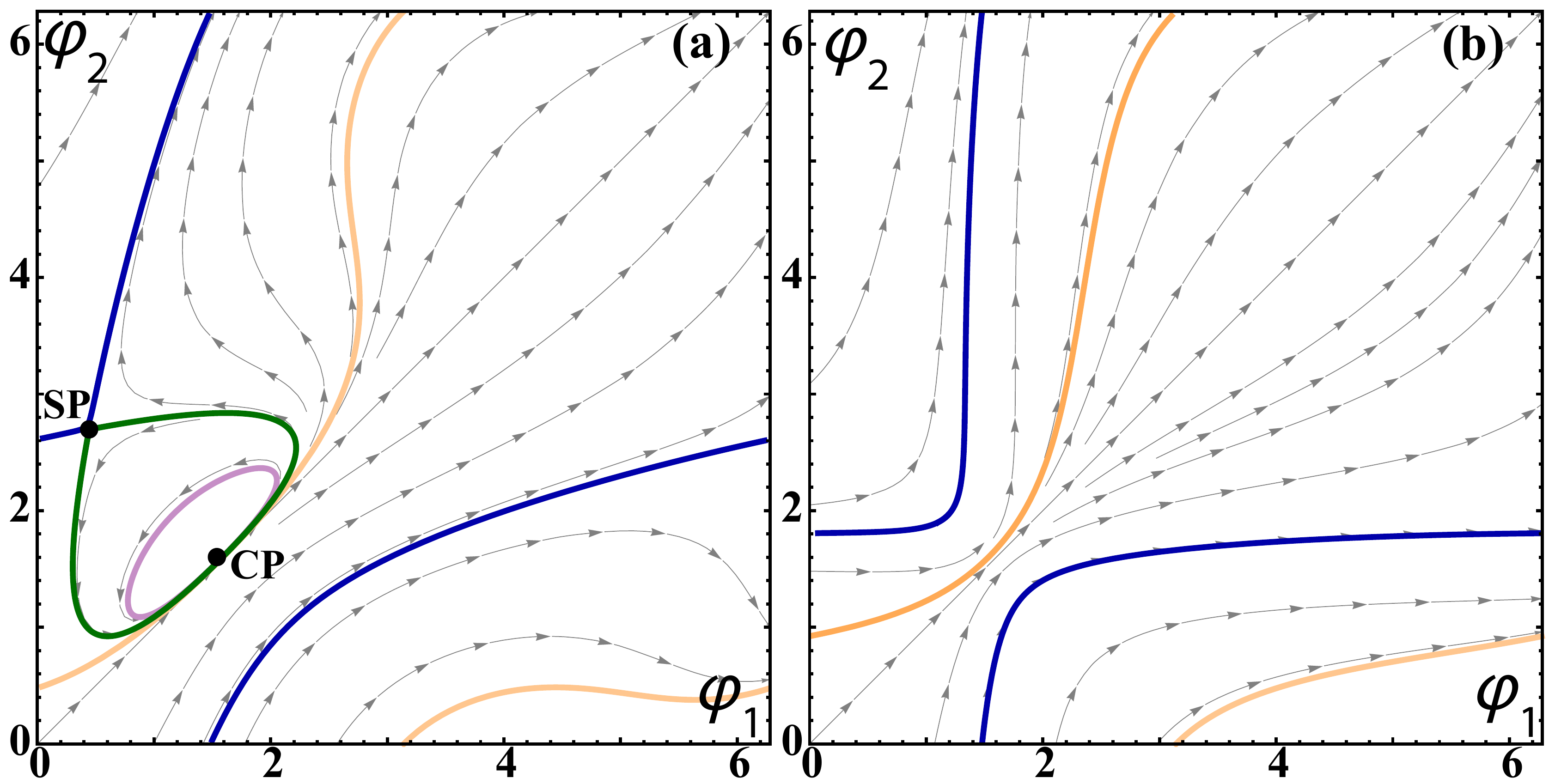}
\caption{Typical dynamics of the fast flow \eqref{eq5} for $I_0=1.05$ below $(\kappa=-0.8)$ and above the saddle-center bifurcation $(\kappa=-0.08)$ are illustrated in (a) and (b), respectively. In (a), the system possesses two unstable fixed points, a saddle (SP) and a center (CP), and exhibits three types of closed orbits: a limit cycle attractor (orange), homoclinic connections to SP (blue and green), and subthreshold  oscillations around the center (purple).
In (b), the system exhibits bistability between two oscillatory states, shown in orange and blue.} \label{fig6}
\vspace{-0.4cm}
\end{figure*}

The structure of the fast flow is organized around the saddle-center bifurcation, which occurs at $\kappa=\kappa_{SC}=-0.1674$. There, the two fixed points get annihilated as a homoclinic orbit associated with the saddle collapses onto the center. To gain a complete picture of the dynamics of the fast flow, we have shown in Figures \ref{fig6}(a) and \ref{fig6}(b) the illustrative examples of the phase portraits and the associated vector fields for $\kappa<\kappa_{SC}$ and $\kappa>\kappa_{SC}$, respectively. For $\kappa\in[-1,\kappa_{SC})$, the fast flow possesses a limit cycle attractor, essentially derived from the local dynamics of the units, cf. the orbit indicated in red in Fig. \ref{fig6}(a). Apart from an attracting periodic orbit, one observes two additional types of closed orbits, namely the homoclinic connections to the saddle point (SP), shown by blue and green, as well as the periodic orbits around the center point (CP), an example of which is indicated in orange. For $\kappa>\kappa_{SC}$, the fast flow exhibits bistability between two oscillatory solutions, such that there is a coexistence of a limit cycle inherited from the local dynamics of units, and the limit cycle associated with the former homoclinic orbits, cf. Fig. \ref{fig6}(b).

\begin{figure*}
\centering
\includegraphics[scale=0.33]{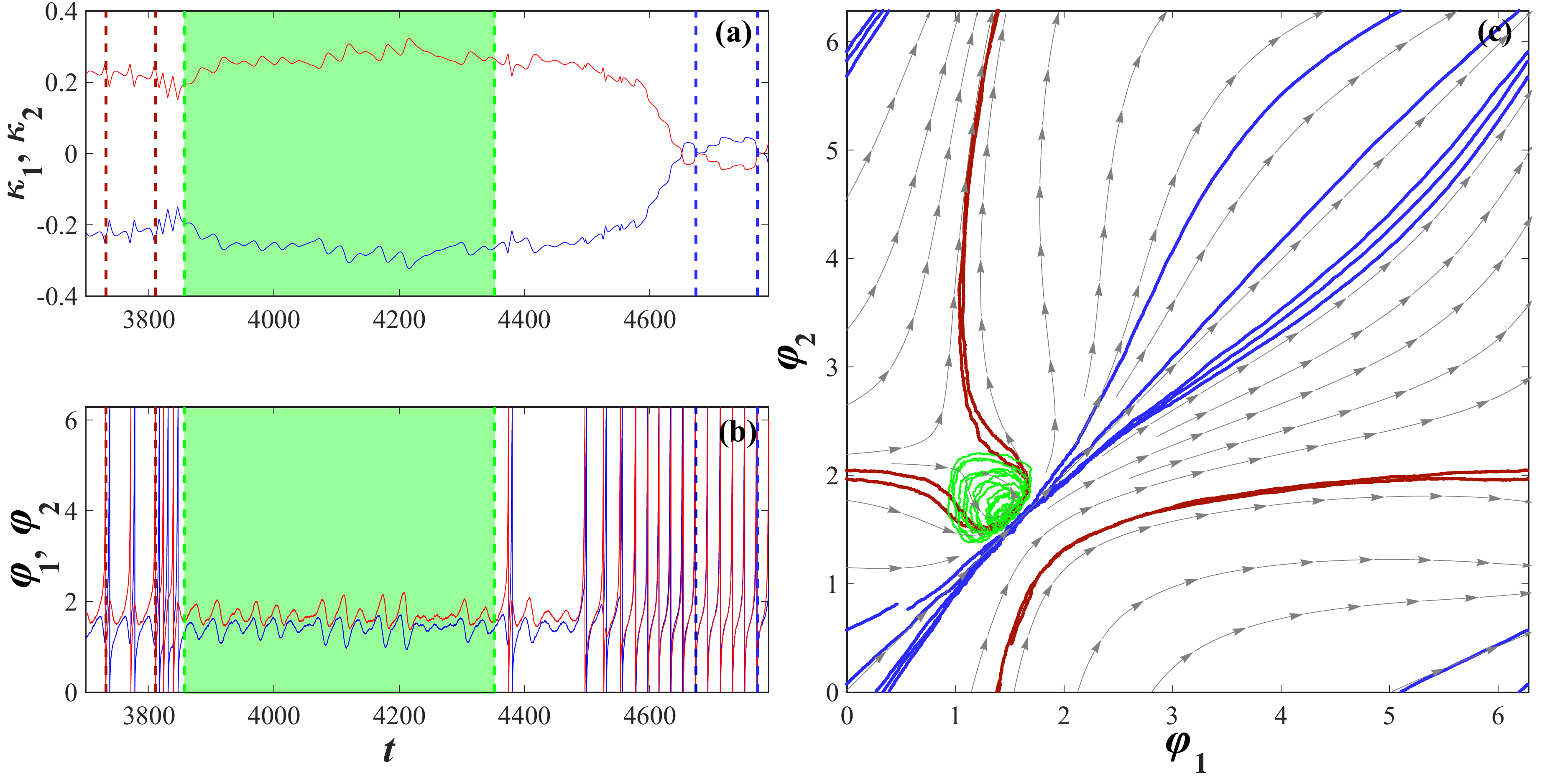}
\caption{(a) and (b) show the time traces of $\kappa_i(t)$ and $\varphi_i(t)$ respectively, with an episode where the system remains in vicinity of an unstable fixed point highlighted in green. The parameters are $I_0=1.05,\varepsilon=0.035, \beta=\pi, D=10^{-4}$. (c) The orbits conforming to the two metastable states characterized by large-amplitude oscillations of phases are shown in red and blue, whereas the subthreshold oscillations are indicated in green. Superimposed is the vector field of the fast flow, corresponding to the limit $\varepsilon\rightarrow 0$.} \label{fig7}
\vspace{-0.3cm}
\end{figure*}
\begin{figure}
\centering
\includegraphics[scale=0.66]{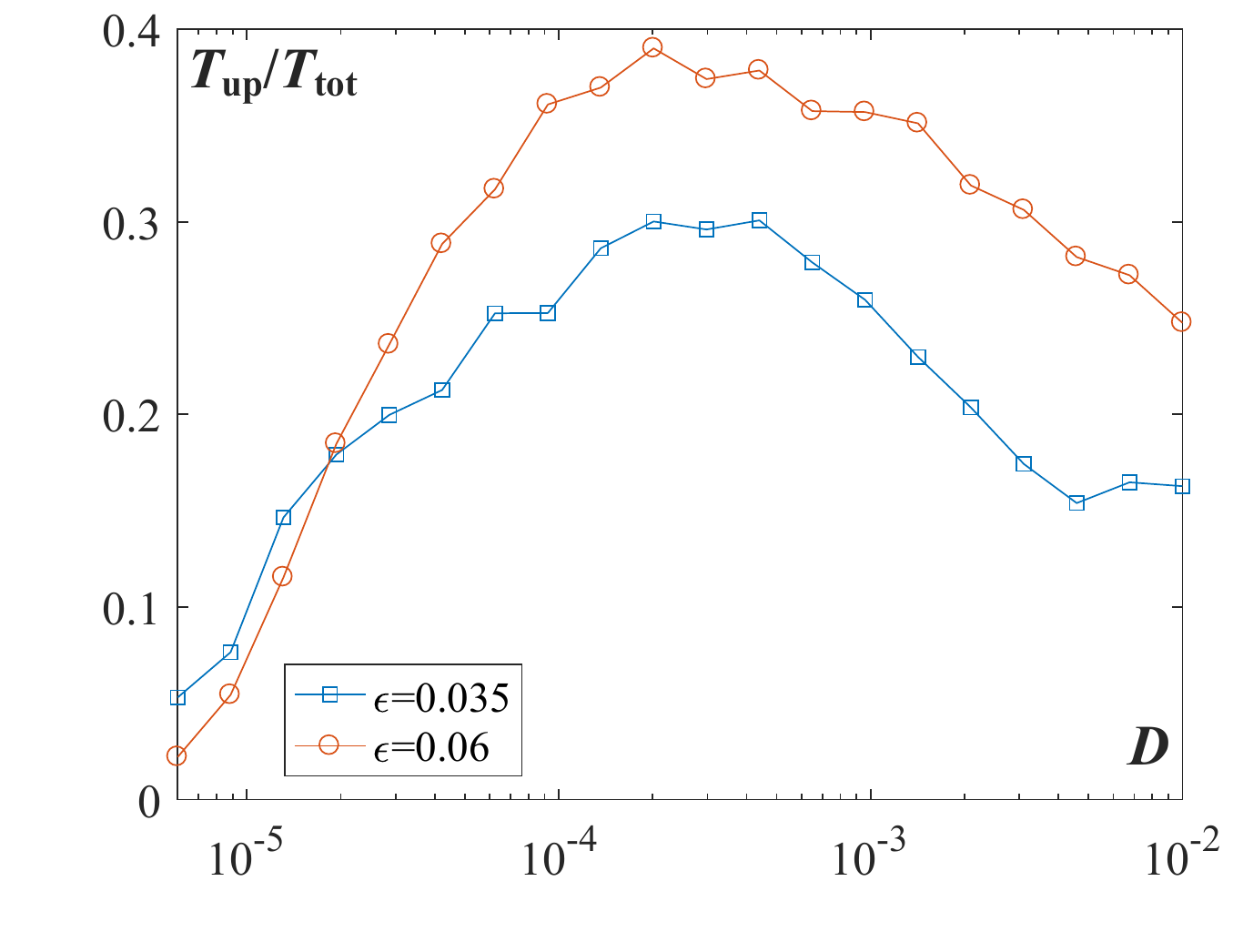}
\caption{Numerically estimated fraction of time spent in vicinity of the unstable fixed point $T_{up}/T_{tot}$ as a function of noise
for $\varepsilon=0.035$ (squares) and $\varepsilon=0.06$ (circles). Note that the positions of the maxima coincide with the corresponding resonant
noise levels from Fig. \ref{fig5}. Remaining system parameters are $I_0=1.05, \beta=\pi$.} \label{fig9}
\vspace{-0.3cm}
\end{figure}

In the presence of noise, the described attractors of the fast flow turn to metastable states. Nevertheless, in contrast to the case of two adaptively coupled excitable units, the slow stochastic fluctuations here do not involve only switching between the metastable states, but also comprise the \emph{subthreshold oscillations} derived from the periodic orbits around the center point. These subthreshold oscillations provide for the trapping effect, which effectively leads to a reduced oscillation frequency. An example of the time series $\kappa_i(t)$ and $\varphi_i(t),i\in\{1,2\}$ obtained for an intermediate $\varepsilon=0.035$ in Fig. \ref{fig7}(a)-(b) indeed shows three characteristic episodes, including visits to two distinct oscillatory metastable states and an extended stay in the vicinity of the center, cf. the stochastic orbits $(\varphi_1(t),\varphi_2(t))$ and the vector field of the fast flow in Fig. \ref{fig7}(c). In the case of finite scale separation, the trapping effect is manifested as the noise-enhanced stability of an unstable fixed point. The prevalence of subthreshold oscillations changes with noise in a non-monotone fashion, see the inset in Fig. \ref{fig7}(c), becoming maximal around the resonant noise level where the frequency dependence on noise exhibits a minimum, cf. Fig. \ref{fig9} and Fig. \ref{fig5}. The fraction of time spent in the metastable state corresponding to subthreshold oscillations has been estimated by the numerical procedure analogous to the one already described in Sec. \ref{sec:m1}.
\vspace{-0.4cm}

%\vspace{-0.4cm}
\section{Two mechanisms of ISR in classical neuronal models} \label{sec:ML}
%\vspace{-0.3cm}

\begin{figure*}
\centering
\includegraphics[scale=0.17]{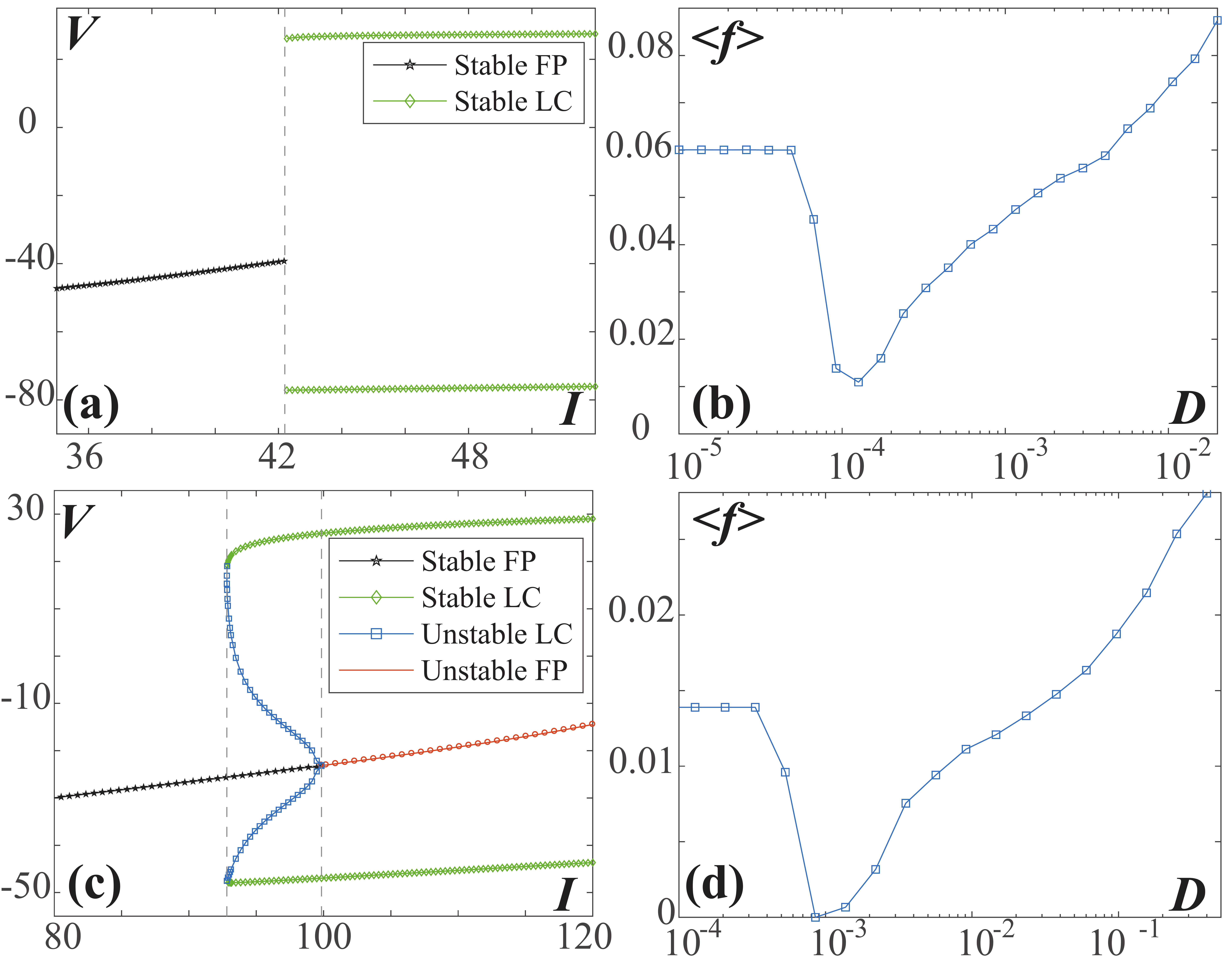}
\caption{(a) Bifurcation diagram showing the dependence of the amplitudes of the membrane potential $V$ on the external bias current $I$ for the version of Morris-Lecar model exhibiting a supercritical Hopf bifurcation. (b) illustrates the $\langle f\rangle(D)$ dependence for the Morris-Lecar neural oscillator in close vicinity of the supercritical Hopf bifurcation. (c) $V(I)$ bifurcation diagram for the setup where the Morris-Lecar model displays a subcritical Hopf bifurcation. (d) Characteristic non-monotone dependence $\langle f\rangle(D)$ for the Morris-Lecar model from (c), with the bifurcation parameter $I=95$ set in the bistable regime. The two sets of parameters putting the Morris-Lecar model in vicinity of a supercritical or a subcritical Hopf bifurcation are specified in the main text.}\label{fig8}
\vspace{-0.3cm}
\end{figure*}
%\vspace{-0.3cm}
So far, we have demonstrated the two paradigmatic scenarios for ISR considering the examples of coupled Type I units, whose local dynamics is close to a SNIPER bifurcation, be it in the excitable or the oscillatory regime. Nevertheless, the onset of ISR and the specific mechanisms of the phenomenon do not depend on the excitability class of local dynamics. In particular, we have recently demonstrated that a single Type II Fitzhugh-Nagumo relaxation oscillator exhibits qualitatively the same form of non-monotone dependence on noise \cite{FOW18}, with the mechanism involving noise-induced subthreshold oscillations that follow the maximal canard of an unstable focus. In that case, it has been established that the trapping effect and the related subthreshold oscillations are triggered due to a phase-sensitive excitability of a limit cycle. Moreover, we have verified that the same model of neuronal dynamics, set to different parameter regimes, may exhibit two different scenarios of ISR. In particular, by an appropriate selection of the system parameters, the Morris-Lecar neuron model
\begin{align}
C\frac{dv}{dt}&=-g_{fast}m(v)(v-E_{Na})-g_{slow}W(v-E_K)-\nonumber \\
&-g_{leak}(v-E_{leak})+I \nonumber \\
\frac{dv}{dt}&=\phi\frac{W_{\infty}(v)-W}{\tau(v)} \nonumber \\
m(v)&=0.5[1+\tanh{(\frac{v-\beta_m}{\gamma_m})}]  \nonumber \\
W_{\infty}(v)&=[ 1+\tanh{(\frac{v-\beta_w}{\gamma_w})}] \nonumber \\
\tau(v)&=1/\cosh{(\frac{v-\beta_w}{2\gamma_w})}, \label{eq8}
\end{align}
where $v$ and $W$ respectively denote the membrane potential and the slow recovery variable, can be placed in vicinity of a supercritical or a subcritical Hopf bifurcation \cite{WWYC11}, with the external bias current $I$ being the bifurcation parameter. In the first case, obtained for $E_{Na}=50\ mV,E_{K}=-100\ mV, E_{leak}=-70\ mV,g_{fast}=20\ mS/cm^2,g_{slow}=20\ mS/cm^2,g_{leak}=2\ mS/cm^2,\phi=0.15,C=2\ \mu F/cm^2,\beta_m=-1.2\ mV,\beta_w=-13\ mV,\gamma_m=18\ mV, \gamma_w=10\ mV$, the model is monostable under the variation of $I$, and the ISR is observed slightly above the Hopf bifurcation ($I=43 \mu A/cm^2$) due to a noise-enhanced stability of an unstable fixed point, cf. Fig. \ref{fig8}(a)-(b). In the second case, conforming to the parameter set $E_{Na}=120\ mV,E_{K}=-84\ mV,E_{leak}=-60\ mV,g_{fast}=4.4\ mS/cm^2,g_{slow}=8\ mS/cm^2,g_{leak}=2\ mS/cm^2,\phi=0.04,C=20\ \mu F/cm^2,\beta_m=-1.2\ mV, \beta_w=2\ mV,\gamma_m=18\ mV,\gamma_w=30\ mV$, the model displays bistability between a limit cycle and a stable equilibrium in a range of $I$ just below the Hopf threshold. There, ISR emerges due to a mechanism based on biased switching, see the bifurcation diagram $V(I)$ in Fig. \ref{fig8}(c) and the dependence of the oscillation frequency on noise for $I=95\ \mu A/cm^2$ in Fig. \ref{fig8}(d).

\vspace{-0.2cm}
\section{Discussion and outlook}
\vspace{-0.3cm}

Considering a model which involves the classical ingredients of neuronal dynamics, such as excitable behavior and coupling plasticity, we have demonstrated two paradigmatic scenarios for inverse stochastic resonance. By one scenario, the phenomenon arises in systems with multistable deterministic dynamics, where at least one of the attractors is a stable equilibrium. Due to the structure of the phase space, and in particular the position of the separatrices, the switching dynamics between the associated metastable states becomes biased at an intermediate noise level, such that the longevity of the quasi-stationary states substantially increases or they may even turn into absorbing states. In the other scenario, an oscillatory system possesses a weakly unstable fixed point, whose stability is enhanced due to the action of noise. The latter results in a trapping effect, such that the system exhibits subthreshold oscillations, whose prevalence is noise-dependent and is found to be maximal at the resonant noise level. Both scenarios involve classical facilitatory effects of noise, such as crossing the separatrices or stochastic mixing across the bifurcation threshold, which should warrant the ubiquity of ISR. In terms of the robustness of the effect, we have demonstrated that the onset of ISR is independent on the excitability class of local dynamics, and moreover, that the same model of neuronal dynamics, depending on the particular parameters, may display two different scenarios for ISR.

Given that ISR has so far been observed at the level of models of individual neurons \cite{TJG09,UCOB13,UTSOB17,FOW18}, motifs of units with neuron-like dynamics \cite{GJT08,BKNPF18} and neural networks \cite{UBT17}, it stands to reason that the phenomenon should be universal to neuronal dynamics, affecting both the emergent oscillations and systems of  coupled oscillators. The explained mechanisms appear to be generic and should be expected in other systems comprised of units with local dynamics poised close to a bifurcation threshold. Inverse stochastic resonance should play important functional roles in neuronal systems, including the reduction of spiking frequency in the absence of neuromodulators, the triggering of stochastic bursting, i.e. of on-off tonic spiking activity, the suppression of pathologically long short-term memories \cite{SM13,UCOB13,UTSOB17,BRHGR16}, and most notably, may contribute to generation of UP-DOWN states, characteristic for spontaneous and induced activity in cortical networks \cite{HMS12,VH13}.

\begin{acknowledgments}
This work was supported by the Ministry of Education, Science and Technological
Development of Republic of Serbia under project No. $171017$. The authors would
also like to thank Matthias Wolfrum and Serhiy Yanchuk for fruitful discussions.
\end{acknowledgments}

\end{document}